\documentclass[conference,10pt]{IEEEtran}
\IEEEoverridecommandlockouts

\usepackage{cite}
\usepackage{graphicx}
\usepackage{amsmath}
\usepackage{array}
\usepackage{mdwmath}
\usepackage{amssymb}
\usepackage{mdwtab}
\usepackage{stfloats}
\usepackage[tight,footnotesize]{subfigure}
\usepackage{amsmath,amsthm}
\usepackage{threeparttable}
\usepackage{color}
\usepackage{url}
\usepackage{algpseudocode}
\usepackage{algorithm}
\usepackage{multicol}
\usepackage{amssymb}
\usepackage{flushend}
\usepackage[bottom=2.8cm,top=2.54cm,left=1.7cm,right=1.7cm]{geometry}

\algrenewcommand{\algorithmicrequire}{\textbf{Input:}}
\algrenewcommand{\algorithmicensure}{\textbf{Output:}}

\newtheorem{remark}{Remark}
\newtheorem{lemma}{Lemma}

\def\BibTeX{{\rm B\kern-.05em{\sc i\kern-.025em b}\kern-.08em
    T\kern-.1667em\lower.7ex\hbox{E}\kern-.125emX}}

\begin{document}

\title{On the Performance of RIS-Aided Spatial Scattering Modulation for mmWave Transmission}

\author{
	\IEEEauthorblockN{
       Xusheng Zhu$^1$, Wen Chen$^1$, Zhendong Li$^1$, Qingqing Wu$^1$, Ziheng Zhang$^1$, Kunlun Wang$^2$, and Jun Li$^3$
}
	\IEEEauthorblockA{1: Department of Electronic Engineering, Shanghai Jiao Tong University, Shanghai, China}
\IEEEauthorblockA{2: School of Communication and Electronic Engineering, East China Normal University, Shanghai, China }
\IEEEauthorblockA{3: School of Electronic and Optical Engineering, Nanjing University of Science Technology, Nanjing, China}
	\IEEEauthorblockA{Email: \{xushengzhu, wenchen, lizhendong, qingqingwu, zhangziheng\}@sjtu.edu.cn\\ klwang@cee.ecnu.edu.cn, jun.li@njust.edu.cn}
}

\markboth{}
{}
\maketitle

\begin{abstract}
In this paper, we investigate a  state-of-the-art reconfigurable intelligent surface (RIS)-assisted spatial scattering modulation (SSM) scheme for millimeter-wave (mmWave) systems, where a more practical scenario that the RIS is near the transmitter while the receiver is far from RIS is considered.  To this end, the line-of-sight (LoS) and non-LoS links are utilized in the transmitter-RIS and RIS-receiver channels, respectively.
By employing the maximum likelihood detector at the receiver, the conditional pairwise error probability (CPEP) expression for the RIS-SSM scheme is derived under the two scenarios that the received beam demodulation is correct or not.
Furthermore, the union upper bound of average bit error probability (ABEP) is obtained based on the CPEP expression.
Finally, the derivation results are exhaustively validated by the Monte Carlo simulations.
\end{abstract}
\begin{IEEEkeywords}
Reconfigurable intelligent surface, spatial scattering modulation, millimeter-wave, average bit error probability.
\end{IEEEkeywords}

\section{Introduction}
Reconfigurable Intelligent Surface (RIS) has risen as a notably promising technique in the advancement of sixth-generation (6G) wireless networks. This is attributed to its distinctive capacity to efficiently modify the propagation environment, all the while diminishing power consumption and hardware cost.
Specifically, RIS is comprised of a large number of low-cost reflective elements arranged in a two-dimensional (2D) planar array configuration \cite{pan2021reconfi}.
Each element of the RIS can be regarded as a reconfigurable scatterer, endowed with the ability to adjust the phase shift to effectively alter the direction of the incident signal \cite{wu2020intelligent}.
By jointly manipulating all elements of the RIS, the reflection signal can be constructively enlarged to improve the signal-to-noise ratio (SNR) or degraded to mitigate interference \cite{li2022joint}.
It is worth noting that RIS has an easy deployment feature, which can improve the delivery of communication services in the desired direction and extend the network coverage \cite{pan2021reconfigurable}. Compared with conventional relays, RIS does not require radio frequency (RF) chains and amplifiers, which can greatly reduce communication energy consumption \cite{wu2019intelligent}.
These advancements in RIS technology make it a promising solution to the challenges of energy consumption and hardware cost in the next-generation communication networks.

\begin{figure*}[t]
\centering
\includegraphics[width=15cm]{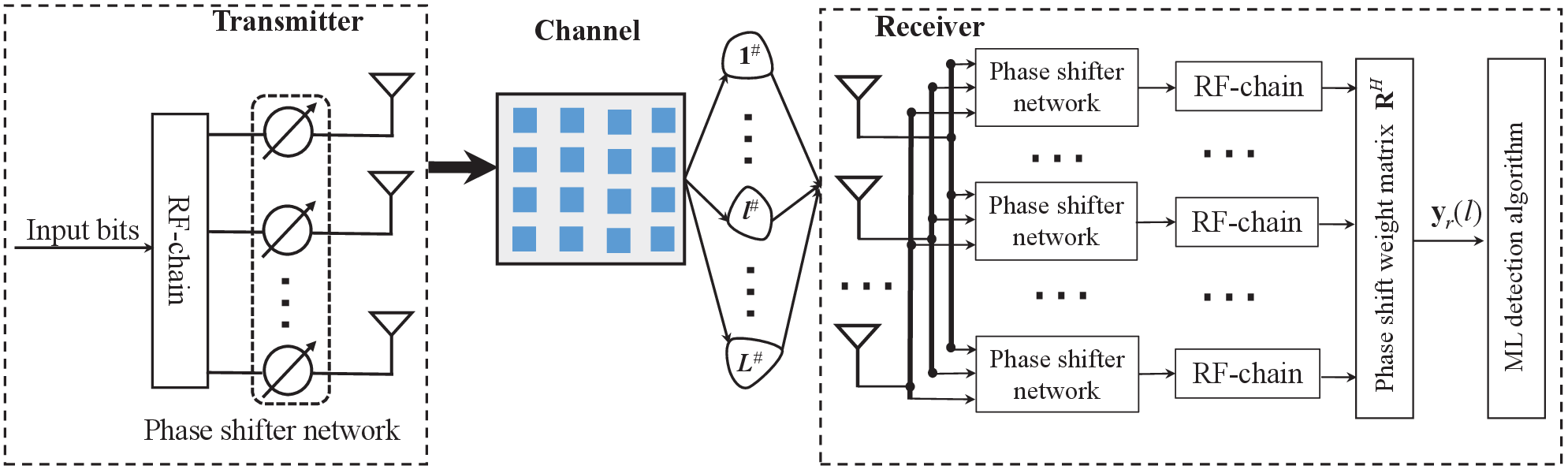}
\caption{\small{System model of the proposed RIS-SSM scheme.}}
\vspace{-10pt}
\label{sys}
\end{figure*}
Index modulation (IM) is a well-established concept in wireless communication that has gained considerable attention in recent decades, which can facilitate the transmission of additional data by utilizing available resources such as transmit antenna, subcarrier, etc. \cite{mao2019nov}.
One of the prominent IM techniques is spatial modulation (SM), which utilizes one RF chain to establish connection with an arbitrary antenna, allowing the spatial location of the antenna to convey additional bits \cite{mes2008spati}.
It is worth noting that SM strikes a favorable balance between energy efficiency and spectral efficiency \cite{renzo2011spatial}.
A recent large-scale measurement campaign was conducted in various indoor environments to investigate the propagation characteristics of SM in real-world scenarios \cite{zhu2022on}.
Moreover, the variant of SM is quadrature SM, which was proposed to improve the overall throughput of conventional SM systems by incorporating additional modulation spatial dimensions \cite{mesleh2015qua}.
Space shift keying (SSK) is a simplified version of SM that relies only on the antenna index to achieve information transmission \cite{jeg2009spac}.
However, in high-frequency bands such as the millimeter-wave (mmWave) range, ensuring satisfactory signal quality becomes challenging when relying solely on the activation of a single antenna during each time slot. To tackle this challenge, spatial scattering modulation (SSM) was introduced in \cite{ding2017spatial}. This approach embraces an analog and hybrid beamforming structure at the transmitter (Tx), enabling the transmission of spatial domain information via scatterer indices.
Similar to SM, SSM also utilizes a single RF chain, empowering the transmitter to concentrate the beam in a specific direction during each time slot.
To improve the spectral efficiency, the quadrature SSM scheme was also investigated in \cite{tu2018iccc}.

Considering these aspects, the potential of IM-based transmission with the help of RIS has been investigated in \cite{easar2020rec,ma2020large,can2020rec,zhu2022re,zhufull2023} while maintaining the inherent advantages of both IM and RIS.
More specifically, Basar first proposed the RIS-SM/SSK schemes and derived the ABEP closed-form expressions, where the RIS aims to transmit a beam at a specific receive antenna to enhance the spectral efficiency based on the spatial domain information \cite{easar2020rec}.
In \cite{can2020rec}, the authors investigated the RIS-SSK scheme in the presence of the channel estimation errors.
To increase the spectral efficiency of the RIS-SSK scheme, \cite{ma2020large} studied the RIS-assisted full-duplex SSK (RIS-FD-SSK) scheme with perfect self-interference (SI) elimination.
Based on this, \cite{zhufull2023} investigated the ABEP performance on the RIS-FD-SSK scheme with the imperfect SI elimination via the Gaussian Chebyshev quadrature approach.
It is worth noting that \cite{easar2020rec,ma2020large,can2020rec,zhufull2023} are all specific to antenna indices, which is not practical in the mmWave band due to the severe path loss.
To fill this gap, \cite{zhu2022re} applied the hybrid beamforming instead of activating different antenna elements in \cite{easar2020rec,ma2020large,can2020rec} by aiming at different scatterers.
It is worth mentioning that in \cite{zhu2022re}, RIS is deployed in the middle of Tx and receiver (Rx), and both Tx-RIS and RIS-Rx use none-line-of-sight (NLoS) link communication. In this paper, we consider another communication scenario, i.e., the RIS is deployed closer to the Tx using the line-of-sight (LoS) path to effectively assist the Tx achieve information transfer and thus significantly mitigate the power consumption and cost of the Tx.
The main contributions of this work are summarized as follows:
1) In this paper, we consider a RIS-assisted SSM mmWave transmission system, where the Tx and RIS are deployed relatively closed positions, they can communicate through the LoS path, while the distance between the RIS and Rx is relatively far from each other, the NLoS paths are considered to convey via the SSM technique in the RIS-Rx channel.
2) A detailed derivation of the conditional pair error probability (CPEP) for RIS-SSM is performed based on maximum likelihood (ML) detection algorithm under two scenarios of correct and incorrect decoding of scatterers.
3) Simulations are utilized to validate the analytical derivations and reveal that the more abundant the scatterers in the environment and the less the number of scatterers involved in the modulation, the better  average bit error probability (ABEP) can be obtained.

\begin{table}[t]
\small
\begin{tabular}{ll}
\hline Notations & Definitions \\
\hline$N_{\mathrm{t}}$ & Number of antenna of the transmit array \\
$N_{\mathrm{r}}$ & Number of antenna of the receive array \\
$n_{\mathrm{t}}$ & Transmit array element index \\
$n_{\mathrm{r}}$ & Receive array element index \\
$L$ & Total number of scatterers \\
$L_s$ & The number of scatterers participating \\& in the modulation \\
$M$ & Symbol domain  phase shift keying/quadrature \\& amplitude modulation (PSK/QAM) modulation order \\
$(\cdot)^{T}$ & The transpose operation \\
$(\cdot)^{{H}}$ & The complex conjugate transpose operation \\
${\rm diag}(\cdot)$ & The diagonal matrix operation \\
$\mathbb{C}^{n\times m}$ & The space of $n\times m$ complex-valued matrices \\
$\mathcal{N}(\cdot,\cdot)$ & The real Gaussian distribution\\
$\mathcal{CN}(\cdot,\cdot)$ & The circularly symmetric complex \\& Gaussian distribution \\
$Q(\cdot)$ & The $Q$-function \\
$\Re\{\cdot\}$ &  The real part of a complex variable \\
$\Pr(\cdot)$ & The probability of an event \\
$P_e$ & The value of CPEP \\
$\sim$ & ``Distributed as"\\
$I_0(\cdot)$  &The modified bessel function \\ & \ \ \ of the first kind of zeroth order \\
$\Gamma(\cdot)$ & The Gamma function \\
$(\cdot)!$ & The factorial operation \\
\hline
\end{tabular}
\end{table}
\section{System Model}
In Fig. \ref{sys}, we consider a RIS-assisted narrowband hybrid mmWave multiple-input multiple-output (MIMO) transmission system model, where the RIS is deployed close to the Tx and can communicate directly with an LoS path, while the RIS is deployed far from the Rx.
\subsection{RIS-SSM Channel Model}
In this paper, we assume that Tx and Rx are equipped with uniform linear array (ULA), while the RIS is composed of the 2D uniform planar array (UPA).

\subsubsection{Tx-RIS Channel}
In this situation, we assume that there exists the strong LoS mmWave MIMO link. As such, the channel between the Tx and RIS can be characterized as
\begin{equation}\label{B1}
\mathbf{B} = \boldsymbol{a}_{ ris}(\phi^r,\varphi^r)\boldsymbol{a}_{t}^H(\theta^t),
\end{equation}
where $\boldsymbol{a}_{t}(\theta^t)$ represents the response of the $N_t$-element ULA at the Tx, which can be given by
\begin{equation}
\boldsymbol{a}_{t}(\theta^t)=\frac{1}{\sqrt{N_t}}
\left[
1,e^{-j2\pi\frac{\delta_t}{\lambda}\sin\theta^t},\cdots,
e^{-j2\pi\frac{\delta_t}{\lambda}(N_t-1)\sin\theta^t}
\right]^T,
\end{equation}
where $\delta_t$ represents the spacing between adjacent elements in the Tx array, $\lambda$ means the carrier wavelength, $\theta^t \in\left[-{\pi},{\pi}\right]$ represents the angle of departure (AoD) of signals from the Tx.
As shown in Fig. \ref{sys}, $\boldsymbol{a}_{ ris}^H(\phi^r,\varphi^r)$ stands for the array response at the RIS in the Tx-RIS channel, which can be evaluated as 
\begin{equation}\label{upar}
\begin{aligned}
&\boldsymbol{a}_{ ris}(\phi^r,\varphi^r) = \frac{1}{\sqrt{N}}
[
1, 
\cdots,e^{-j2\pi\frac{\delta_{ris}}{\lambda}(n_{\rm h}\sin\phi^r\cos\varphi^r+
n_{\rm v}\cos\phi^r)},\\&
\cdots,e^{-j2\pi\frac{\delta_{ris}}{\lambda}((N_{\rm h}-1)\sin\phi^r\cos\varphi^r+(N_{\rm v}-1)\cos\phi^r)}
]^T,
\end{aligned}
\end{equation}
where $N_{\rm h}$ and $N_{\rm v}$ are the number of rows and columns of RIS, respectively.
$n_{\rm h}$ and $n_{\rm v}$ represent the corresponding row index and column index, respectively.
$\delta_{ris}$ denotes the spacing between adjacent reflective RIS elements.
$\phi^r \in \left[-{\pi},{\pi}\right]$ represents the elevation angle of arrival (AoA) of incoming signals to the RIS and $\varphi^r \in \left[-{\pi},{\pi}\right]$ denotes the azimuth AoA of signals to the RIS.

\subsubsection{RIS-Rx Channel}
Due to the small wavelength of the mmWave, it has limited ability to diffract on encountering obstacles.
In this way, mmWave channels exhibit a sparse multipath structure and are typically characterized by the Saleh-Valenzuela channel model.
It is worth noting that the SSM technology working in the mmWave band is adopted to complete the information transmission of the RIS-Rx channel.
Accordingly, the channel from the RIS to Rx can be expressed as
\begin{equation}\label{F}
\mathbf{F}= \sum_{l=1}^Lh_l \boldsymbol{a}_r(\theta_{l}^r)
\boldsymbol{a}_{ris}^H(\phi_l^t,\varphi_l^t),
\end{equation}
where $L$ is the total number of signal paths between the RIS and the Rx.
$\phi_l^t \in \left[-{\pi},{\pi}\right]$ and $\varphi_l^t \in \left[-{\pi},{\pi}\right]$ denote the azimuth and elevation AoDs associated with the RIS, respectively. $\theta_l^r \in \left[-{\pi},{\pi}\right]$ represents the AoA at the Rx side, $h_l$ is the complex channel gain.
$\boldsymbol{a}_{r}(\theta^r)$ denotes the normalized array response vectors associated with the Rx.
Hence, the ULA with $N_r$ antennas at the Rx side can be given as
\begin{equation}
\boldsymbol{a}_{r}(\theta_l^r)=\frac{1}{\sqrt{N_r}}
\left[
1,e^{-j2\pi\frac{\delta_r}{\lambda}\sin\theta_l^r},\cdots,
e^{-j2\pi\frac{\delta_r}{\lambda}(N_r-1)\sin\theta_l^r}
\right]^T,
\end{equation}
where $\delta_r$ and $\lambda$ are the antenna spacing and the signal wavelength.
Besides, the normalized array response transmitted from the RIS can be given by
\begin{equation}\label{upat}
\begin{aligned}
&\boldsymbol{a}_{ ris}(\phi_l^t,\varphi_l^t) =\frac{1}{\sqrt{N}}
[
1, \cdots,e^{-j2\pi\frac{\delta_{ris}}{\lambda}(n_{\rm h}\sin\phi_l^t\cos\varphi_l^t+n_{\rm v}\cos\phi_l^t)},\\&\cdots,
e^{-j2\pi\frac{\delta_{ris}}{\lambda}((N_{\rm h}-1)\sin\phi_l^t\cos\varphi_l^t+(N_{\rm v}-1)\cos\phi_l^t)}
]^T.
\end{aligned}
\end{equation}
\subsubsection{Composite Channel }
The transmitted signal propagates through the Tx-RIS-Rx channel to reach Rx, where the RIS consists of $N$ passive reflecting elements.
Each of them is independent of each other without interference.
To characterize the performance limit of RIS, we assume the magnitude of each reflecting element equals one.
As this point, the reflection matrix of the RIS can be formulated as
\begin{equation}\label{psi}
\boldsymbol{\Psi}={\rm diag}(e^{j\psi_1},\cdots,e^{j\psi_n},\cdots,e^{j\psi_N}) \in \mathbb{C}^{N\times N}.
\end{equation}
Considering Eqs. (\ref{B1}), (\ref{F}), and (\ref{psi}), the composite channel can be calculated  as
\begin{equation}\label{h11}
\begin{aligned}
\mathbf{H}&= \mathbf{F} \boldsymbol{\Psi} \mathbf{B}= \sum_{l=1}^Lh_l \boldsymbol{a}_r(\theta_{l}^r)
\boldsymbol{a}_{ris}^H(\phi_l^t,\varphi_l^t) \boldsymbol{\Psi}
\boldsymbol{a}_{ ris}(\phi^r,\varphi^r)\boldsymbol{a}_{t}^H(\theta^t).
\end{aligned}
\end{equation}
Let us define $\beta_l\overset{\Delta}{=}h_l\boldsymbol{a}_{ris}^H(\phi_l^t,\varphi_l^t)$,
$\zeta^r = \arg [\boldsymbol{a}_{ ris}^H(\phi^r,\varphi^r)]$, and
$\zeta^t_l = \arg [\boldsymbol{a}_{ris}(\phi_l^t,\varphi_l^t)]$.
Based on this, the phase of composite channel is
$
{\Xi}
= \frac{1}{{N}}\sum_{n=1}^Ne^{j(\psi_n-\zeta^t_l-\zeta^r)}.
$
By tuning the phase shifts of RIS as $\psi_n=\zeta^t_l+\zeta^r$, we have $\sum_{n=1}^Ne^{j(\psi_n-\zeta^t_l-\zeta^r)} = N$.
Hence, the Eq. (\ref{h11}) can be further expressed as
\begin{equation}\label{eq12}
\begin{aligned}
\mathbf{H}
&= \sum\nolimits_{l=1}^Lh_l \boldsymbol{a}_r(\theta_{l}^r)
\boldsymbol{a}_{t}^H(\theta^t).
\end{aligned}
\end{equation}
\begin{lemma}
As the number of antennas of $N_t$ and $N_r$ is large enough, we can argue that the beams directed between the different scatterers are orthogonal to each other. Thus, we have
\begin{equation}\label{ortho}
\begin{aligned}
\boldsymbol{a}_t^H(\theta_l^t)\boldsymbol{a}_t(\theta_{l'}^t) = \delta(l-l'), \ \ \boldsymbol{a}_r^H(\theta_l^r)\boldsymbol{a}_r(\theta_{l'}^r) = \delta(l-l').
\end{aligned}
\end{equation}
\end{lemma}
{\emph{Proof:}}
 Please refer Appendix A for the detailed proof. $\hfill\blacksquare$

\subsection{RIS-SSM Transmission}
In Fig. \ref{sys}, we consider that the RIS is deployed close to the Tx, where the Tx and RIS communicate through the LoS link. Among the RIS-Rx channels, it is difficult to guarantee the existence of a stable LoS path due to the existence of abundant scatterers. Consequently, the RIS-Rx can realize information transfer through the NLoS paths.
In the proposed RIS-SSM scheme, the SSM technique is employed in the RIS-Rx channel, where the RIS can accurately emit the reflecting beam towards the objective scatterer in each time slot based on the perfect channel state information (CSI).
To enhance the reliability of the transmission, we select $L_s$ candidate scatterers in descending order of amplitude gain out of the $L$ scatterers in the RIS-Rx channel, i.e., $|h_1|>|h_2|>\cdots>|h_{L}|$.
As such, the received information at the Rx side can be given as
\begin{equation}
\mathbf{y} = \sqrt{{P_s}}\mathbf{H}\boldsymbol{a}_{t}(\theta^t) s_m + {\mathbf{n}},
\end{equation}
where $P_s$ denotes the transmit power, and $\mathbf{n}$ represents additive Gaussian white noise vector following $\mathcal{CN}(0,N_0\mathbf{I}_{N_r})$.

\subsection{RIS-SSM Detection}

The maximum ratio combing technique is applied to achieve reception of the signal through the $l$-th scatterer at the Rx side.
To distinguish that the received beam is coming from a specific scatterer, the Rx is equipped with no less than $L_s$ RF chains.
Notice that each RF chain has a set of phase shifters to monitor a specific scatterer.
In this respect, it is feasible to monitor the $L_s$ scattered beams in real time, thus enabling the recovery of the signal in the spatial domain.
Here, the combined signal after the RF chains can be given by
\begin{equation}
\mathbf{y}_r=
[\boldsymbol{a}_r^H\left(\theta_1^r\right),\dots,\boldsymbol{a}_r^H\left(\theta_{L_{\mathrm{s}}}^r\right)]^T
\mathbf{y}.
\end{equation}
Next, the ML detector is performed to jointly detects the activated scatterer index and the modulated symbols, which can be given by
\begin{equation}\label{ml}
{[{\hat{l}}, {\hat{m}}]}=
\mathop{\mathop{\mathop\mathbf{{\arg\min}} _{{m \in\{1, \ldots, M\}}}}_{{l \in\{1, \ldots, L_s\}}}}
|\mathbf{y}_{{r}}\left(l\right)-\boldsymbol{a}_r^H(\theta_{ l}^r)\mathbf{H}{\boldsymbol{a}_t(\theta^t)}\sqrt{P_s}s_m |^{2},
\end{equation}
where $\mathbf{y}_r(l)$ denotes the $l$-th element of $\mathbf{y}_r$,
$\hat l$ and $\hat s_m$ denotes the detected index of scatterer and symbol, respectively.

\section{Performance Analysis}
In this section,  we  derive the CPEP for union detection of the selected scatterer index $l$ and the transmitted symbol $m$.
In particular, the mathematical derivation process can be divided into Rx correctly detected and falsely detected scatterers.
\subsection{$\hat{l} = l$}
In this case, the detection of transmission direction is correct, whereas the error comes from the transmitted symbol, i.e., $s_m \neq s_{\hat m}$.
Herein, the CPEP can be calculated as
\begin{equation}\label{pr1}
\begin{aligned}
P_e=\Pr&\left(|\mathbf{y}_r (l) - \boldsymbol{a}_r^H(\theta_{ l}^r)\mathbf{H}\boldsymbol{a}_t(\theta^t)\sqrt{P_s}s_m |^2\right.\\
&>\left.|\mathbf{y}_r ({ l}) - \boldsymbol{a}_r^H(\theta_{\hat l}^r)\mathbf{H}\boldsymbol{a}_t(\theta^t)\sqrt{P_s}s_{\hat m} |^2\right).
\end{aligned}
\end{equation}
Combining Eqs. (\ref{ortho}) and (\ref{pr1}), we can get
\begin{equation}\label{rorth}
\begin{cases}
\mathbf{y}_{{r}}\left(l\right)
= \mathbf{R}^H  \mathbf{y}\left(l,m\right)
=
\sqrt{P_s}h_{l}s_m+\boldsymbol{a}_r^H \left(\theta_{l}^r\right)\mathbf{n},\\
|\mathbf{y}_r ({l}) - \boldsymbol{a}_r^H(\theta_{ l}^r)\mathbf{H}{\boldsymbol{a}_t(\theta^t)}\sqrt{P_s}s_m |^2
=|\boldsymbol{a}_r^H(\theta_l^r)\mathbf{n}|^2,\\
\left|\mathbf{y}_r ({\hat l}) - \boldsymbol{a}_r^H(\theta_{\hat l}^r)\mathbf{H}{\boldsymbol{a}_t(\theta^t)}\sqrt{P_s}s_{\hat m} \right|^2\\=
\left|\sqrt{P_s}h_{l}\left(s_m-s_{\hat m}\right)+\boldsymbol{a}_r^H \left(\theta_{l}^r\right)\mathbf{n}\right|^2.
\end{cases}
\end{equation}
Substituting Eq. (\ref{rorth}) into Eq. (\ref{pr1}), the CPEP can be evaluated as
\begin{equation}\label{sdfg}
\begin{aligned}
P_e=&\Pr(-|\sqrt{P_s}h_{l}\left(s_m-s_{\hat m}\right)|^2\\
&>|2\boldsymbol{a}_r^H \left(\theta_{l}^r\right)\mathbf{n}\sqrt{P_s}h_{l}\left(s_m-s_{\hat m}\right)|^2)\\
=&\Pr(-|\sqrt{P_s}h_{l}\left(s_m-s_{\hat m}\right)|^2\\&
-2\Re\{n_r\sqrt{P_s}h_{l}\left(s_m-s_{\hat m}\right)\}>0)\\
=&\Pr\left(G>0\right),
\end{aligned}
\end{equation}
where 
$G$ follows $\mathcal{N}(\mu_G,\sigma_G^2)$ with
$\mu_G=-|\sqrt{P_s}h_{l}\left(s_m-s_{\hat m}\right)|^2$ and
$\sigma_G^2 = |2N_0\sqrt{P_s}h_{l}\left(s_m-s_{\hat m}\right)|^2$.
Based on this, we have
\begin{equation}\label{qqq}
\begin{aligned}
&P_e=Q\left(-{\mu_G}/{\sigma_G}\right)=Q\left(\sqrt{\frac{\rho |h_l|^2|s_m - s_{\hat m}|^2}{2}}\right),
\end{aligned}
\end{equation}
where $\rho = {P_s}/{N_0}$ denotes average SNR.
\subsection{$\hat{l} \neq l$}
In this case, the transmission direction is detected incorrectly, and the erroneous bits originate from both the beam and signal components.
Consequently, the CPEP can be written as
\begin{equation}\label{lnelpr}
\begin{aligned}
P_e
=&\Pr\left(|\mathbf{y}_r (l) - \boldsymbol{a}_r^H(\theta_{ l}^r)\mathbf{H}\boldsymbol{a}_t(\theta^t)\sqrt{P_s}s_m |^2\right.\\&
>\left.|\mathbf{y}_r ({\hat l}) - \boldsymbol{a}_r^H(\theta_{\hat l}^r)\mathbf{H}\boldsymbol{a}_t(\theta^t)\sqrt{P_s}s_{\hat m} |^2\right).
\end{aligned}
\end{equation}
Depending on Eqs. (\ref{ortho}) and (\ref{lnelpr}), we can obtain
\begin{equation}\label{rorthh}
\begin{cases}
\mathbf{y}_{{r}}(\hat l)
= \mathbf{R}^H  \mathbf{y}(\hat l,\hat m)
=
\boldsymbol{a}_r^H(\theta_{\hat l}^r)\mathbf{n},\\
|\mathbf{y}_r ({l}) - \boldsymbol{a}_r^H(\theta_{ l}^r)\mathbf{H}{\boldsymbol{a}_t(\theta^t)}\sqrt{P_s}s_m |^2
=|\boldsymbol{a}_r^H(\theta_l^r)\mathbf{n}|^2,\\
|\mathbf{y}_r ({\hat l}) \!-\! \boldsymbol{a}_r^H(\theta_{\hat l}^r)\mathbf{H}{\boldsymbol{a}_t(\theta^t)}\sqrt{P_s}s_{\hat m}|^2\!=\!
|\boldsymbol{a}_r^H(\theta_{\hat{l}}^r) \mathbf{n}\!-\!\sqrt{P_s}h_{\hat l}s_{\hat{m}}|^2.
\end{cases}
\end{equation}
Combining  Eqs. (\ref{lnelpr}) and (\ref{rorthh}), we have
\begin{equation}\label{eq22}
\begin{aligned}
P_e
=&\Pr\left(|\boldsymbol{a}_r^H(\theta_l^r)\mathbf{n}|^2
>|\boldsymbol{a}_r^H(\theta_{\hat{l}}^r) \mathbf{n}-\sqrt{P_s}h_{\hat l}s_{\hat{m}}|^2\right).
\end{aligned}
\end{equation}
For brevity, let us define
$\varrho_1 = |\boldsymbol{a}_r^H(\theta_l^r)\mathbf{n}|^2$ and
$\varrho_2 =|\boldsymbol{a}_r^H(\theta_{\hat{l}}^r) \mathbf{n}-\sqrt{P_s}h_{\hat l}s_{\hat{m}}|^2$,
where $\boldsymbol{a}_r^H(\theta_l^r)\mathbf{n} \sim \mathcal{CN}(0,N_0)$ and $(\boldsymbol{a}_r^H(\theta_{\hat{l}}^r) \mathbf{n}-\sqrt{P_s}h_{\hat l}s_{\hat{m}} ) \sim \mathcal{CN}(-\sqrt{P_s}h_{\hat l}s_{\hat{m}},N_0)$.
Note that ${\varrho_1}$ denotes the central Chi-squared random variable with two degrees of freedom (DoFs), and ${\varrho_2}$ represents the non-central Chi-squared random variable with two DoFs.
To facilitate the solution, we make ${x_1 = \frac{\varrho_1}{N_0/2}}$ and
$x_2 = \frac{\varrho_2}{N_0/2}$.
As such, the Eq. (\ref{eq22}) can be calculated as
\begin{equation}\label{np1}
\begin{aligned}
P_e=&\int_0^\infty\int_{x_2}^\infty f_1(x_1) f_2(x_2)dx_1 d x_2,
\end{aligned}
\end{equation}
where the central Chi-square distribution can be described as
\begin{equation}\label{np2}
f_1(x_1)=\frac{1}{2}\exp\left(-\frac{x_1}{2}\right).
\end{equation}
Substituting Eq. (\ref{np2}) into Eq. (\ref{np1}), the $P_e$ can be updated as
\begin{equation}\label{np4}
\begin{aligned}
P_e
=&\int_0^\infty f_2(x_2) \exp\left(-\frac{x_2}{2}\right) d x_2.
\end{aligned}
\end{equation}
Besides, the non-central Chi-square distribution can be computed by \cite{kopka}
\begin{equation}\label{f2x2}
f_2(x_2)=\frac{1}{2}{\exp\left(-\frac{x_2+\tau}{2}\right)}I_0(\sqrt{kx}),
\end{equation}
where $\tau = 2\rho|h_{\hat{l}}|^2|s_{\hat{m}}|^2$ and $I_0(\cdot)$ can be given as
\begin{equation} \label{I01}
\begin{aligned}
I_0(\sqrt{kx})&=\sum_{k=0}^\infty\frac{\left(\frac{kx}{4}\right)^k}{k!\Gamma(k+1)}\overset{(a)}{=}\sum_{k=0}^\infty\frac{\left(\frac{kx}{4}\right)^k}{k!k!},
\end{aligned}
\end{equation}
where $(a)$ stands for
$
\Gamma(k+1)=\int_0^\infty t^{k}e^{-x}dx=k!.
$
Combining Eqs. (\ref{f2x2}) and (\ref{I01}), we have
\begin{equation}\label{npep3}
f_2(x_2)=\frac{ x_2^k}{2}{\exp\left(-\frac{x_2}{2}\right)}{\exp\left(-\frac{\tau}{2}\right)}\sum\limits_{k=0}^\infty
{\left(\frac{\tau}{4}\right)^k}\frac{1}{k!k!}.
\end{equation}
Next, we substitute Eq. (\ref{npep3}) into Eq. (\ref{np4}), then the
CPEP in Eq. (\ref{np4}) can be updated as
\begin{equation}\label{npep4}
\begin{aligned}
P_e
=&\frac{1}{2}{\exp\left(-\frac{\tau}{2}\right)}\sum\limits_{k=0}^\infty
{\left(\frac{\tau}{4}\right)^k}\frac{1}{k!k!}\int_0^\infty x_2^k \exp\left(-{x_2}\right) d x_2.\\
\end{aligned}
\end{equation}
Recall that the Gamma function, the CPEP can be reformulated as
\begin{equation}\label{npep5}
\begin{aligned}
P_e
=&\frac{1}{2}{\exp\left(-\frac{\tau}{2}\right)}\sum\limits_{k=0}^\infty
{\left(\frac{\tau}{4}\right)^k}\frac{1}{k!}.\\
\end{aligned}
\end{equation}
By utilizing $\exp(x)=\sum_{k=0}^\infty\frac{x^k}{k!}$, the CPEP can be obtained as
\begin{equation}\label{cpep2}
P_e=\frac{1}{2}{\exp\left(-\frac{\rho|h_{\hat{l}}|^2|s_{\hat{m}}|^2}{2}\right)}.
\end{equation}

\subsection{ABEP}
Based on the CPEP, the union upper bound of ABEP can be written as
\begin{equation}\label{abep}
\begin{aligned}
{\rm ABEP} \leq &\frac{1}{ML_s\log_2({ML_s})}\sum_{l=1}^L\sum_{m=1}^M\sum_{\hat l=1}^L\sum_{\hat m=1}^M\bar{P}_e\\&\times N(\left[{l}, {m}\right] \rightarrow[\hat{l}, \hat{m}] ),
\end{aligned}
\end{equation}
where $N(\left[{l}, {m}\right] \rightarrow[\hat{l}, \hat{m}])$ stands for the Hamming distance between the binary expressions of the symbols $\hat l$ to $l$ and $\hat m$ to $m$.

\section{Numerical and Analytical Results}
In this section, simulation results and analytical derivations of the RIS-SSM scheme are provided.
For the simulation, the channel realization is generated by $1 \times 10^6$.
According to \cite{tu2018iccc}, let us set $N_r=32$, $N_t=32$, $\delta_t=\lambda/2$, and $\delta_r=\lambda/2$.
Besides, the derivation results analyze the impact of various RIS-SSM system configurations.

\begin{figure}[t]
\centering
\includegraphics[width=5cm]{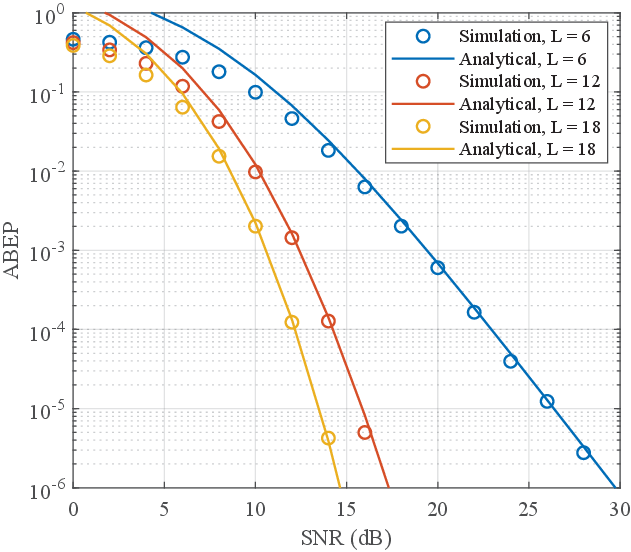}
\caption{\small{Impact of the total number of scatterers on ABEP. }}
\label{fig1}
\end{figure}

\begin{figure}[t]
\centering
\includegraphics[width=5cm]{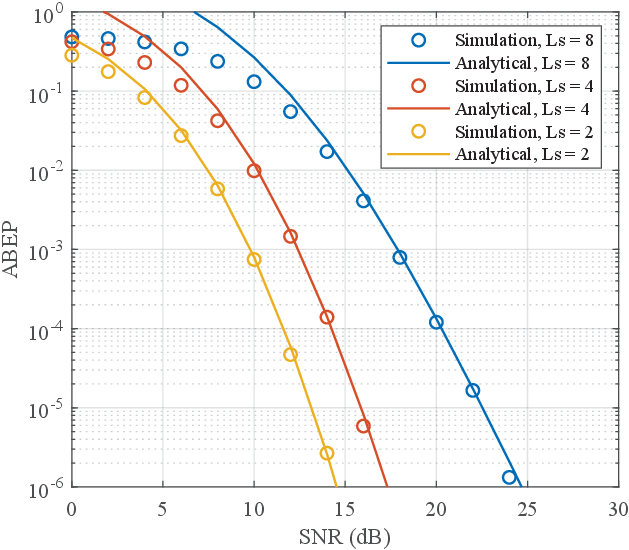}
\caption{\small{Impact of modulating scatterer order on ABEP. }}
\label{fig2}
\end{figure}

\begin{figure}[t]
\centering
\includegraphics[width=5cm]{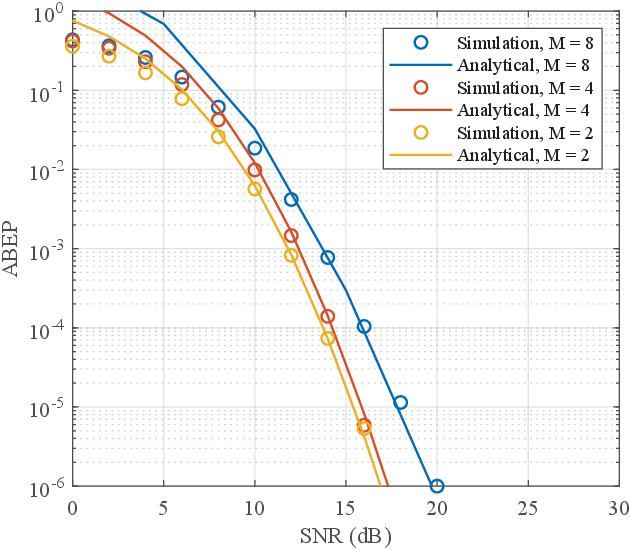}
\caption{\small{ABEP with different modulation orders in the symbol domain. }}
\label{fig3}
\end{figure}

In Fig. \ref{fig1}, we present the ABEP performance of the proposed RIS-SSM scheme with the different, where $L_s = 4$ and QPSK modulation is adopted. It is seen from Fig. \ref{fig1} that with increasing SNR, ABEP decreases rapidly when the total number of scatterers is 12 and 18, while ABEP decreases slightly more slowly as the total number of scatterers is 6.
This is because among the $L$ scatterers in the channel, we select $L_s$ scatterers in descending order of amplitude to participate in the modulation of the proposed RIS-SSM scheme.
In other words, when the scatterers in the environment are more enriched, the quality of the scatterers selected by the descending order algorithm is higher, thus leading to a better ABEP performance for the proposed RIS-SSM scheme.

In Fig. \ref{fig2}, we investigate the impact of the order of the scatterers participating in the modulation on the ABEP performance in the RIS-SSM scheme.
As depicted in Fig. \ref{fig2}, the ABEP performance improves as the modulation order $L_s$ decreases from 8 to 2.
To be specific, when ABEP = $1\times 10^{-6}$, the RIS-SSM system with $L_s=2$ requires 3 dB and 10 dB less SNRs than $L_s = 4$ and $L_s = 8$, respectively.
This is because that when the value of $L_s$ is small, the RIS steers its beam towards fewer scatterers in each transmission slot. This results in a lower modulation order in the spatial domain, that is, signal detection needs to be reduced by traversing the complexity of detection, which results in a lower probability of decoding errors.

In Fig. \ref{fig3}, we compare the ABEP performance of the RIS-SSM scheme for $L=12$ and $L_s=4$ cases, where the $L_s$ with the highest gain are selected in descending order of $L$ scatterer gain to be modulated.
The symbol modulation order $M$ is set to be 2, 4, and 8, respectively.
As shown in Fig. \ref{fig3}, the proposed RIS-SSM scheme with BPSK outperforms its counterpart with QPSK and 8PSK.
Specifically, at ABEP = $1\times10^{-6}$, RIS-SSM with 8PSK and QPSK requires 2 dB and 0.5 dB more SNR than in the case of BPSK.
This is because the Euclidean distance between the adjacent constellation points becomes shorter as the modulation order is higher in the normalized symbol domain constellation diagram, thus leading to an increase in the probability of false detection.

\section{Conclusion}
In this paper, we presented a novel  RIS-SSM scheme for the mmWave MIMO communication, where the RIS is deployed closer to the Tx to transmit information with an LoS link, while the RIS is farther away from the Rx to transmit information with NLoS paths.
Based on this architecture, we apply the ML detector to derive the CPEP expression under the received beam is detected correctly and incorrectly, respectively.
Combing the CPEP and Hamming distance, we obtain the union upper bound of the ABEP.
The Monte Carlo simulation provides an exhaustive verification of the analytical derivation results.
It demonstrated that the more abundant the scatterers in the environment, the better the ABEP performance of the proposed RIS-SSM system. In addition, the system performance can be improved by employing low-order modulation in the spatial and symbol domains.

\section{Acknowledgement}
This work is supported by National key project 2020YFB1807700, NSFC 62071296, Shanghai 22JC1404000, and PKX2021-D02.

\begin{appendices}
\section{}
For the large-scale ULA, the array response vectors of $N_t$-antenna can be expressed as
\begin{equation}\label{ula3}
\begin{aligned}
&\boldsymbol{a}^H(\theta_{l_1})\boldsymbol{a}(\theta_{l_2})
=  \frac{1}{\sqrt{N_t}}[1,e^{-j2\pi\phi_{l_1}},\cdots,e^{-j2\pi\phi_{l_1}(N_t-1)}]\\
&\times \frac{1}{\sqrt{N_t}}[1,e^{j2\pi\phi_{l_2}},\cdots,e^{j2\pi\phi_{l_2}(N_t-1)}]^T\\
=&\frac{e^{j\pi(\phi_{l_2}-\phi_{l_1})N_t}}{N_te^{j\pi(\phi_{l_2}-\phi_{l_1})}}\times\frac{e^{-j\pi(\phi_{l_2}-\phi_{l_1})N_t}-e^{j\pi(\phi_{l_2}-\phi_{l_1})N_t}}{e^{-j\pi(\phi_{l_2}-\phi_{l_1})}-e^{j\pi(\phi_{l_2}-\phi_{l_1})}}\\
=&\frac{1}{N_t}\frac{\sin\left(\pi(\phi_{l_2}-\phi_{l_1})N_t\right)}{\sin\left(\pi(\phi_{l_2}-\phi_{l_1})\right)}e^{j\pi(\phi_{l_2}-\phi_{l_1})(N_t-1)}.
\end{aligned}
\end{equation}
Afterwards, we divide the subsequent proof into $l_1 = l_2$ and $l_1 \neq l_2$ cases.

\begin{remark}
If $l_1 = l_2$, it is easy to have $\boldsymbol{a}^H(\theta_{l_1})\boldsymbol{a}(\theta_{l_2})=1$.
\end{remark}
\begin{remark} If $l_1 \neq l_2$, we define a function as $f(x) = \left|\frac{\sin(N_tx)}{N_t\sin(x)}\right|$. Substituting $f(x)$ into $\alpha_{l_1,l_2}$, we have
\begin{equation}
\begin{aligned}
&f(\pi(\phi_{l_2}-\phi_{l_1}))=\left|\frac{\sin(N_t\pi(\phi_{l_2}-\phi_{l_1}))}{N_t\sin(\pi(\phi_{l_2}-\phi_{l_1}))}\right|\\
&\approx\left|\frac{\sin(N_t\pi(\phi_{l_2}-\phi_{l_1}))}{N_t\pi(\phi_{l_2}-\phi_{l_1})}\right|
\leq \left|\frac{1}{N_t\pi(\phi_{l_2}-\phi_{l_1})}\right|.
\end{aligned}
\end{equation}
Let us define
$\Delta = \phi_{l_2}-\phi_{l_1}$, then $f(x_0)$ can be rewritten as
\begin{equation}
\begin{aligned}
f(\pi(\phi_{l_2}-\phi_{l_1}))
&=\left|\frac{2}{N_t\pi \Delta}\right|.
\end{aligned}
\end{equation}
When  $N_t\to \infty$, we have
\begin{equation}
\begin{aligned}
\lim\limits_{N_t \to \infty}f(\pi(\phi_{l_2}-\phi_{l_1}))
&=\lim\limits_{N_t \to \infty}\left|\frac{2}{N_t\pi \Delta}\right|= 0.
\end{aligned}
\end{equation}
\end{remark}
At this time, the proof of {\bf Lemma 1} is completed.
\end{appendices}

\end{document}